\documentclass[letterpaper,12pt]{article}
\usepackage[hmargin=1in,bmargin=1.5in,tmargin=1.2in]{geometry}
\pdfoutput=1
\usepackage{Sweave}
\usepackage{color}
\usepackage{booktabs}
\usepackage{colortbl}
\usepackage{dcolumn}
\usepackage{multirow}
\usepackage{rotating}
\usepackage{tabularx}

\usepackage{tikz}
\usetikzlibrary{calc}
\pgfdeclarelayer{background}
\pgfdeclarelayer{foreground}
\pgfsetlayers{background,main,foreground}\definecolor{colorM1}{RGB}{241,235,230}
 \definecolor{colorM2}{RGB}{211,189,168}
 \definecolor{colorM3}{RGB}{184,152,124}
 \definecolor{colorM4}{RGB}{124,108,92}
 \definecolor{colorM5}{RGB}{88,48,8}
 \definecolor{colorA1}{RGB}{195,229,208}
 \definecolor{colorA2}{RGB}{164,229,188}
 \definecolor{colorA3}{RGB}{86,156,112}
 \definecolor{colorA4}{RGB}{67,90,75}
 \definecolor{colorA5}{RGB}{6,64,27}
 \definecolor{colorB1}{RGB}{192,205,233}
 \definecolor{colorB2}{RGB}{124,147,197}
 \definecolor{colorB3}{RGB}{67,93,148}
 \definecolor{colorB4}{RGB}{53,66,93}
 \definecolor{colorB5}{RGB}{10,31,75}
\usepackage{amssymb,amsmath}
\newcommand{\bx}{\boldsymbol x}
\newcommand{\btheta}{\boldsymbol \theta}
\newcommand{\myvar}{\mbox{Var}}
\newcommand{\myexp}{\mbox{E}}
\newcommand{\norm}[1]{\lVert#1\rVert}
\newcommand{\VignetteIndexEntry}[1]{}

\usepackage{fancyhdr}
\begin{document}
\thispagestyle{plain}\begin{center}{\LARGE Marginal Likelihood Estimation via Arrogance Sampling }\\

\vspace{1em}{\large By  Benedict Escoto }\\

\end{center}
\VignetteIndexEntry{Technical paper explaining and justifying technique}
\begin{abstract}
  This paper describes a method for estimating the marginal likelihood
  or Bayes factors of Bayesian models using non-parametric importance
  sampling (``arrogance sampling'').  This method can also be used to
  compute the normalizing constant of probability distributions.
  Because the required inputs are samples from the distribution to be
  normalized and the scaled density at those samples, this method may
  be a convenient replacement for the harmonic mean estimator.  The
  method has been implemented in the open source R package
  \texttt{margLikArrogance}.
\end{abstract}

\section{Introduction}

When a Bayesian evaluates two competing models or theories, $T_1$ and
$T_2$, having observed a vector of observations $\bx$, Bayes' Theorem
determines the posterior ratio of the models' probabilities:

\begin{equation} \label{bayes factor}
  \frac{p(T_1|\bx)}{p(T_2|\bx)} = \frac{p(\bx|T_1)}{p(\bx|T_2)} \frac{p(T_1)}{p(T_2)}.
  \end{equation}

\noindent The quantity $\frac{p(\bx|T_1)}{p(\bx|T_2)}$ is called a
\emph{Bayes factor} and the quantities $p(\bx|T_1)$ and $p(\bx|T_2)$
are called the theories' \emph{marginal likelihoods}.

The types of Bayesian models considered in this paper have a fixed
finite number of parameters, each with their own probability function.
If $\btheta$ are parameters for a model $T$, then

\begin{equation} \label{main integral}
  p(\bx|T) = \int p(\bx|\btheta, T) p(\btheta|T) \, d\btheta
  = \int p(\bx \wedge \btheta|T) \, d\btheta
\end{equation}

\noindent Unfortunately, this integral is difficult to compute in
practice.  The purpose of this paper is to describe one method for
estimating it.

Evaluating integral (\ref{main integral}) is sometimes called the
problem of computing normalizing constants.  The following formula
shows how $p(\bx|T)$ is a normalizing constant.

\begin{equation}
  \label{norm constant}
  p(\btheta|\bx, T) = \frac{p(\btheta \wedge \bx|T)}{p(\bx|T)}
  \end{equation}

\noindent Thus the marginal likelihood $p(\bx|T)$ is also the
normalizing constant of the posterior parameter distribution
$p(\btheta|\bx, T)$ assuming we are given the density $p(\btheta
\wedge \bx|T)$ which is often easy to compute in Bayesian models.
Furthermore, Bayesian statisticians typically produce samples from the
posterior parameter distribution $p(\btheta|\bx, T)$ even when not
concerned with theory choice.  In these case, computing the marginal
likelihood is equivalent to computing the normalizing constant of a
distribution from which samples and the scaled density at these
samples are available. The method described in this paper takes this
approach.

\section{Review of Literature}
\label{lit review}

Given how basic (\ref{bayes factor}) is, it is perhaps surprising that
there is no easy and definitive way of applying it, even for simple
models.  Furthermore, as the dimensionality and complexity of
probability distributions increase, the difficulty of approximation
also increases.  The following three techniques for computing bayes
factors or marginal likelihoods are important but will not be
mentioned further here.

\begin{enumerate}
  \item Analytic asymptotic approximations such as Laplace's method,
    see for instance Kass and Raftery (1995),
  \item Bridge sampling/path sampling/thermodynamic integration
    (Gelman and Meng, 1998), and
  \item Chib's MCMC approximation (Chib, 1995; Chib and Jeliazkov, 2005).
\end{enumerate}

\noindent Kass and Raftery (1995) is a popular overview of the earlier
literature on Bayes factor computation.  All these methods can be very
successful in the right circumstances, and can often handle problems
too complex for the method described here.  However, the method of
this paper may still be useful due to its convenience.

The rest of section \ref{lit review} describes three approaches that
are relevant to this paper.

\subsection{Importance Sampling}
\label{imp sampling}

Importance sampling is a technique for reducing the variance of monte
carlo integration.  This section will note some general facts; see
Owen and Zhou (1998) for more information.

Suppose we are trying to compute the (possibly multidimensional)
integral $I$ of a well-behaved function $f(\btheta)$.  Then

\begin{equation*}
  I = \int f(\btheta) \, d\btheta =
  \int \frac{f(\btheta)}{g(\btheta)} g(\btheta) \, d(\btheta)
\end{equation*}

\noindent so if $g(\btheta)$ is a probability density function and
$\btheta_i$ are independent samples from it, then 

\begin{equation}
  \label{imp approx}
  I = \myexp_g[f(\btheta)/g(\btheta)] \approx \frac{1}{n} \sum_{i=1}^n \frac{f(\btheta_i)}{g(\btheta_i)} = I_n.
\end{equation}

\noindent $I_n$ is an unbiased approximation to $I$ and by the central
limit theorem will tend to a normal distribution.  It has variance

\begin{equation}
  \label{imp var}
  \myvar[I_n] = \frac{1}{n} \int \left(\frac{f(\btheta)}{g(\btheta)} - I\right)^2 g(\btheta)\, d\btheta = 
  \frac{1}{n} \int \frac{(f(\btheta) - Ig(\btheta))^2}{g(\btheta)}\, d\btheta
\end{equation}

\noindent Sometimes $f$ is called the \emph{target} and $g$ is called
the \emph{proposal} distribution.

Assuming that $f$ is non-negative, then minimum variance (of $0!$) is
achieved when $g = f / I$---in other words when $g$ is just the
normalized version of $f$.  This cannot be done in practice because
normalizing $f$ requires knowing the quantity $I$ that we wanted to
approximate; however (\ref{imp var}) is still important because it
means that the more similar the proposal is to the target, the better
our estimator $I_n$ becomes.  In particular, $f$ must go to 0 faster
than $g$ or the estimator will have infinite variance.

To summarize this section:

\begin{enumerate}
  \item Importance sampling is a monte carlo integration technique
    which evaluates the target using samples from a proposal
    distribution.
  \item The estimator is unbiased, normally distributed, and its
    variance (if not 0 or infinity) decreases as $O(n^{-1})$ (using
    big-$O$ notation).
  \item The closer the proposal is to the target, the better the
    estimator.  The proposal also needs to have longer tails than the
    target.
\end{enumerate}

\subsection{Nonparametric Importance Sampling}

A difficulty with importance sampling is that it is often difficult to
choose a proposal distribution $g$.  Not enough is known about $f$ to
choose an optimal distribution, and if a bad distribution is chosen
the result can have large or even infinite variance.  One approach to
the selection of proposal $g$ is to use non-parametric techniques to
build $g$ from samples of $f$.  I call this class of techniques
self-importance sampling, or \textbf{arrogance sampling} for short,
because they attempt to sample $f$ from itself without using any
external information.  (And also isn't it a bit arrogant to try to
evaluate a complex, multidimensional integral using only the values at
a few points?)  The method of this paper falls into this class and
particularly deserves the name because the target and proposal (when
they are both non-zero) have exactly the same values up to a
multiplicative constant.

Two papers which apply nonparametric importance sampling to the
problem of marginal likelihood computation (or computation of
normalizing constants) are Zhang (1996) and Neddermeyer (2009).
Although both authors apply their methods to more general situations,
here I will use the framework suggested by (\ref{norm constant}) and
assume that we can compute $p(\btheta \wedge \bx|T)$ for arbitrary
$\btheta$ and also that we can sample from the posterior parameter
distribution $p(\btheta|\bx, T)$.  The goal is to estimate the
normalizing constant, the marginal likelihood $p(\bx|T)$.

Zhang's approach is to build the proposal $g$ using traditional kernel
density estimation.  $m$ samples are first drawn from $p(\btheta|\bx,
T)$ and used to construct $g$.  Then $n$ samples are drawn from $g$
and used to evaluate $p(\bx|T)$ as in traditional importance sampling.
This approach is quite intuitive because kernel estimation is a
popular way of approximating an unknown function.  Zhang proves that
the variance of his estimator decreases as $O(m^\frac{-4}{4+d}n^{-1})$
where $d$ is the dimensionality of $\btheta$, compared to $O(n^{-1})$
for standard (parametric) importance sampling.

There were, however, a few issues with Zhang's method:

\begin{enumerate}
  \item A kernel density estimate is equal to 0 at points far from the
    points the kernel estimator was built on.  This is a problem
    because importance sampling requires the proposal to have longer
    tails than the target.  This fact forces Zhang to make the
    restrictive assumption that $p(\btheta|\bx,T)$ has compact
    support.
  \item It is hard to compute the optimal kernel bandwidth.  Zhang
    recommends using a plug-in estimator because the function
    $p(\btheta \wedge \bx|T)$ is available, which is unusual for
    kernel estimation problems.  Still, bandwidth selection appears to
    require significant additional analysis.
  \item Finally, although the variance may decrease as
    $O(m^\frac{-4}{4+d}n^{-1})$ as $m$ increases, the difficulty of
    computing $g(\btheta)$ also increases with $m$, because it
    requires searching through the $m$ basis points to find all the
    points close to $\btheta$.  In multiple dimensions, this problem
    is not trivial and may outweigh the $O(m^\frac{-4}{4+d})$ speedup
    (in the worst case, practical evaluation of $g(\btheta)$ at a
    single point may be $O(m)$).  See Zlochin and Baram (2002) for
    some discussion of these issues.
\end{enumerate}

\noindent Neddermeyer (2009) uses a similar approach to Zhang and also
achieves a variance of $O(m^\frac{-4}{4+d}n^{-1})$.  It improves on
Zhang's approach in two ways relevant to this paper:

\begin{enumerate}
  \item The support of $p(\btheta|\bx,T)$ is not required to be
    compact.
  \item Instead of using kernel density estimators, linear blend
    frequency polynomials (LBFPs) are used instead.  LBFPs are
    basically histograms whose density is interpolated between
    adjacent bins.  As a result, the computation of $g(\btheta)$
    requires only finding which bin $\btheta$ is in, and looking up
    the histogram value at that and adjacent bins ($2^d$ bins in
    total).
\end{enumerate}

As we will see in section \ref{my technique}, the arrogance sampling
described in this paper is similar to the methods of Zhang and
Neddermeyer.

\subsection{Harmonic Mean Estimator}

The harmonic mean estimator is a simple and notorious method for
calculating marginal likelihoods.  It is a kind of importance
sampling, except the proposal $g$ is actually the distribution
$p(\btheta|\bx, T) = p(\btheta \wedge \bx|T) / p(\bx|T)$ to be
normalized and the target $f$ is the known distribution
$p(\btheta|T)$.  Then if $\btheta_i$ are samples from
$p(\btheta|x,T)$, we apparently have

\begin{equation*}
  1 \approx \frac{1}{n} \sum_{i=1}^{n}
  \frac{p(\btheta_i|T)}{p(\btheta_i|\bx, T)} = \frac{1}{n} \sum_{i=1}^{n}
  \frac{p(\btheta_i|T)}{p(\bx|\btheta_i, T)p(\btheta_i|T) /
    p(\bx|T)}
  = \frac{1}{n} \sum_{i=1}^{n} \frac{1}{p(\bx|\btheta_i, T) / p(\bx|T)}
\end{equation*}

\noindent hence

\begin{equation} \label{hme}
  p(\bx|T) \stackrel{?}{\approx} \left(\frac{1}{n} \sum_{i=1}^{n}
  \frac{1}{p(\bx|\btheta_i, T)} \right)^{-1}
\end{equation}

Two advantages of the harmonic mean estimator are that it is simple to
compute and only depends on samples from $p(\btheta|x, T)$ and the
likelihood $p(\bx|\btheta,T)$ at those samples.  The main drawback of
the harmonic mean estimator is that it doesn't work---as mentioned
earlier the importance sampling proposal distribution needs to have
longer tails than the target.  In this case the target $p(\btheta|T)$
typically has longer tails than the proposal $p(\btheta|\bx, T)$ and
thus (\ref{hme}) has infinite variance.  Despite not working, the
harmonic mean estimator continues to be popular (Neal, 2008).

\section{Description of Technique}
\label{my technique}

This paper's arrogance sampling technique is a simple method that
applies the nonparametric importance techniques of Zhang and
Neddermeyer in an attempt to develop a method almost as convenient as
the harmonic mean estimator.

The only required inputs are samples $\btheta_i$ from $p(\btheta|\bx,
T)$ and the values $p(\btheta_i \wedge \bx|T) =
p(x|\btheta_i,T)p(\btheta_i|T)$.  This is similar to the harmonic mean
estimator, but perhaps slightly less convenient because $p(\btheta_i
\wedge \bx|T)$ is required instead of $p(\bx|\btheta_i,T)$.

There are two basic steps:

\begin{enumerate}
  \item Take $m$ samples from $p(\btheta|\bx, T)$ and using
    modified histogram density estimation, construct probability
    density function $f(\btheta)$.
  \item With $n$ more samples from $p(\btheta|\bx, T)$, estimate
    $1/p(\bx|T)$ via importance sampling with target $f$ and proposal
    $p(\btheta|\bx, T)$.
\end{enumerate}

\noindent These steps are described in more detail below.

\subsection{Construction of the Histogram}

Of the $N$ total samples $\btheta_i$ from $p(\btheta|\bx, T)$, the
first $m$ will be used to make a histogram.  The optimal choice of $m$
will be discussed below, but in practice this seems difficult to
determine.  An arbitrary rule of $\mbox{min}(0.2 N, 2 \sqrt{N})$ can
be used in practice.

With a traditional histogram, the only available information is the
location of the sampled points.  In this case we also know the
(scaled) heights $p(\btheta \wedge \bx|T)$ at each sampled point.  We
can use this extra information to improve the fit.

Our ``arrogant'' histogram $f$ is constructed the same as a regular
histogram, except the bin heights are not determined by the number of
points in each bin, but rather by the minimum density over all points
in the bin.  If a bin contains no sampled points, then $f(\btheta) =
0$ for $\btheta$ in that bin.  Then $f$ is normalized so that $\int
f(\btheta) \,d\btheta = 1$.

To determine our bin width, we can simply and somewhat arbitrarily set
our bin width $h$ so that the histogram is positive for 50\% of the
sampled points from the distribution $p(\btheta|\bx, T)$.  To
approximate $h$, we can use a small number of samples (say, 40) from
$p(\btheta|\bx, T)$ and set $h$ so that $f(\btheta) > 0$ for exactly
half of these samples.

\begin{figure}\begin{center}\includegraphics[width=15.5cm,height=18cm]{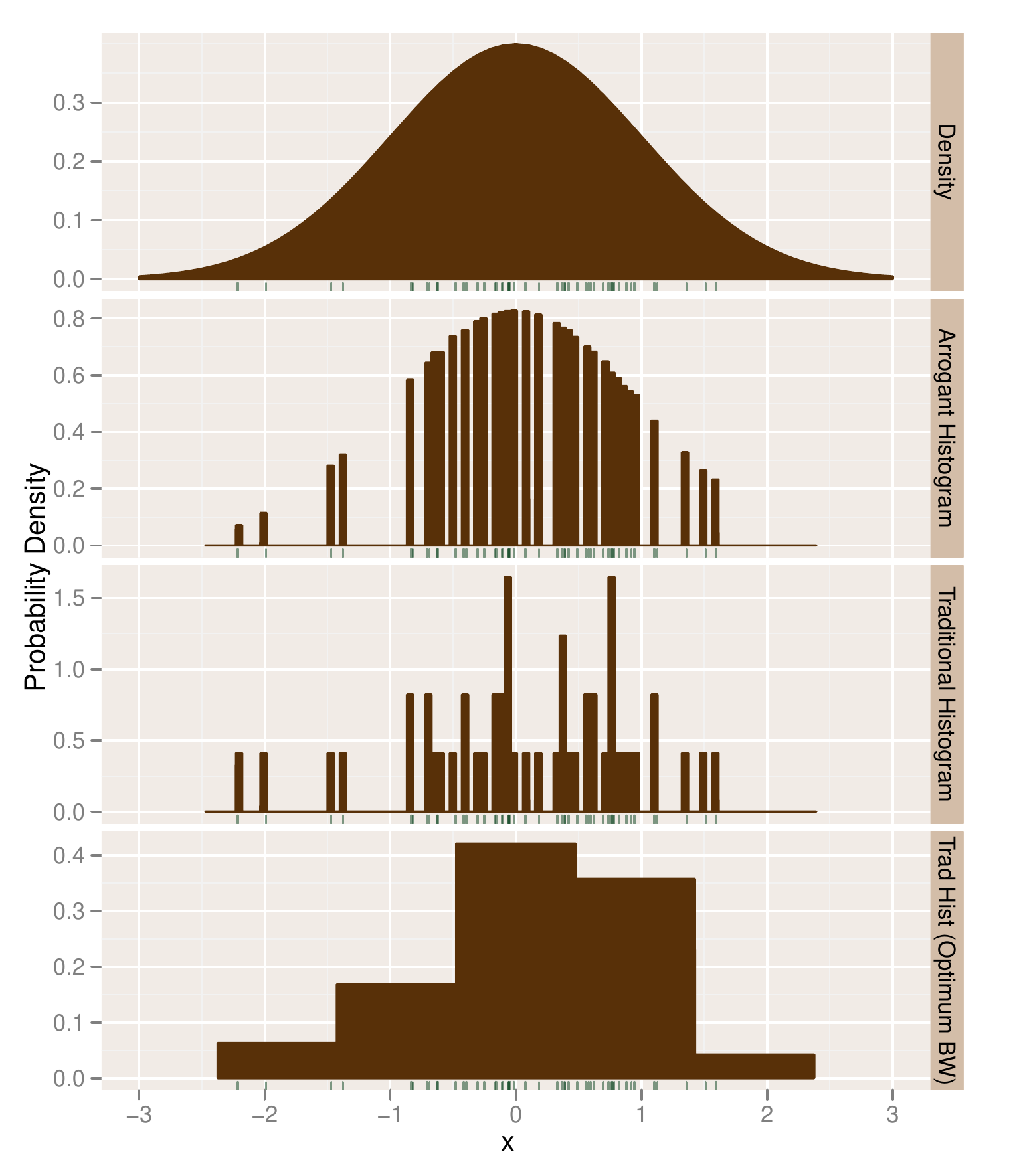}\caption{Histogram Comparison}\label{histograms}\end{center}\end{figure}

Figure \ref{histograms} compares the traditional and new histograms
for a one dimensional normal distribution based on 50 samples.  The
green rug lines indicate the $50$ sampled points which are the same
for all.  The arrogant histogram's bin width is chosen as above.  The
traditional histogram's optimal bin width was determined by Scott's
rule to minimize mean squared error.  As the figure shows, the
modified histogram is much smoother for a given bin width, so a
smaller bin width can be used.  On the other hand, $f$ will either
equal 0 or have about twice the original density at each point, while
the traditional histogram's density is numerically close to the
original density.

\subsection{Importance Sampling}

The remaining $n = N - m - 40$ sampled points can be used for
importance sampling.  Using equation (\ref{imp approx}) with histogram $f$ as
our target and $p(\btheta|\bx, T)$ as the proposal, we have

\begin{equation*}
  1 \approx I_n = \frac{1}{n} \sum_{i=1}^n \frac{f(\btheta_i)}{p(\btheta_i|\bx, T)}
  = \frac{1}{n} \sum_{i=1}^n \frac{f(\btheta_i)}{p(\btheta_i \wedge \bx|T) / p(\bx|T)}
\end{equation*}

\noindent hence

\begin{equation} \label{arrogance}
  p(\bx|T) \approx p(\bx|T) / I_n = \left( \frac{1}{n} \sum_{i=1}^n
  \frac{f(\btheta_i)}{p(\btheta_i \wedge \bx|T)} \right)^{-1} = A_n
\end{equation}

\noindent To underscore the self-important/arrogant nature of
this approximation $A_n$, we can rewrite (\ref{arrogance}) as

\begin{equation*}
  p(\bx|T) \approx H \left( \frac{1}{n} \sum_{i=1}^n
    \frac{\mbox{min}\{p(\btheta_j \wedge \bx|T): \btheta_j \mbox{ and } \btheta_j \mbox{ are in the same bin}\}}{p(\btheta_i \wedge \bx|T)} \right)^{-1}
\end{equation*}

\noindent where $H$ is the histogram normalizing constant.  This
equation shows that all the values in the numerator and the
denominator of our importance sampling are from the same distribution 
$p(\btheta \wedge \bx|T)$.

Note that the histogram $f$ is the target of the importance sampling
and $p(\btheta \wedge \bx|T)$ is the proposal.  This is backwards from
the usual scheme where the unknown distribution is the target and the
known distribution is the proposal.  Instead here the unknown
distribution is the proposal, as in the harmonic mean estimator (see
Robert and Wraith (2009) for another example of this.)

As in section \ref{imp sampling}, our approximation of $p(\bx|T)^{-1}$
tends to a normal distribution as $n \to \infty$ by the central limit
theorem.  This fact can be used to estimate a confidence interval
around $p(\bx|T)$.

\section{Validity of Method}

This section will investigate the performance of the method.  First,
note that this method is just an implementation of importance
sampling, so $A_n^{-1}$ should converge to $p(\bx|T)^{-1}$ with finite
variance as long as the proposal density $p(\btheta|\bx, T)$ exists
and is finite and positive on the compact region where the target
histogram density is positive.

To calculate the speed of convergence we will use equation (\ref{imp
  var}) where $f$ is the histogram, $g(\btheta) = p(\btheta|\bx, T)$,
and $I = 1$ because the histogram has been normalized.  Unless
otherwise noted, we will assume below that $g: \mathbb{R}^d
\rightarrow \mathbb{R}$ is finite, twice differentiable and positive,
and that $\int \frac{\norm{\nabla \cdot g(\btheta)}^2}{g(\btheta)}
d\btheta$ is finite.

\subsection{Histogram Bin Width}

One important issue will be how quickly the $d$-dimensional
histogram's selected bin width $h$ goes to 0 as the number of samples
$m \rightarrow \infty$.  This section will only offer an intuitive
argument.  For any $m$, the histogram will enclose about the same
probability ($\frac{1}{2}$) and will have about the same average
density in a fixed region.  Each bin has volume $h^d$, so if $l$ is
the number of bins then $lh^d = O(1)$ and $h \propto l^{-d}$.

Furthermore, the distribution of the sampled points converges to the
actual distribution $g(\btheta)$.  If $m > O(l)$, an unbounded number
of sampled points would end up in each bin.  If $m < O(l)$, then some
bins would have no points in them.  Neither of these is possible
because exactly one sampled point is necessary to establish each bin.
Thus $m \propto l$ and  $h \propto m^{-d}$.

\subsection{Conditional Variance}

Before estimating the convergence rate of $A_n$ we will prove
something about the conditional variance of importance sampling.  Let
$A = \{\btheta: f(\btheta) > 0\}$, $\mathbf{1}_A$ be the characteristic
function of $A$, and $q = \int_A g(\btheta) \,d\btheta$.  Define

\begin{equation*}
  g_A(\btheta) = \left\{
  \begin{array}{cl}
    g(\btheta)/q & \mbox{ if } \btheta \in A\\
    0 & \mbox{ otherwise }
  \end{array} \right.
\end{equation*}

\noindent Then $g_A$ is the density of $g$ conditional on $f > 0$.
Define $\mbox{Var}_A$ and $\mbox{E}_A$ to mean the variance and
expectation conditional on $f(\btheta) > 0$.  Thus

\begin{eqnarray*}
  \mbox{Var}(f(\btheta)/g(\btheta)) &=& \mbox{Var}(\mbox{E}(f(\btheta)
  / g(\btheta) | \mathbf{1}_A)) + \mbox{E}(\mbox{Var}(f(\btheta) / g(\btheta) | \mathbf{1}_A))\\
  &=& \mbox{Var}\left(
  \begin{array}{cl}
    \mbox{E}_A(f(\btheta) / g(\btheta)) & \mbox{ if } \btheta \in A\\
    0 & \mbox{ otherwise}\\
  \end{array}\right)\\
  &&  + \,\mbox{E}\left(
  \begin{array}{cl}
    \mbox{Var}_A(f(\btheta) / g(\btheta)) & \mbox{ if } \btheta \in A\\
    0 & \mbox{ otherwise}\\
  \end{array}\right)\\
  &=& \mbox{Var}\left(
  \begin{array}{cl}
    1/q & \mbox{ if } \btheta \in A\\
    0 & \mbox{ otherwise}\\
  \end{array}\right) + q\mbox{Var}_A(f(\btheta) / g(\btheta))\\
  &=& (1/q)^2q(1-q) + \frac{1}{q}\mbox{Var}_A(f(\btheta) / q g(\btheta))\\
  &=& \frac{1-q}{q} + \frac{1}{q}\mbox{Var}_A(f(\btheta) / g_A(\btheta))
\end{eqnarray*}

We will assume below that $q=\frac{1}{2}$, so that

\begin{equation}
  \label{I var}
  \mbox{Var}(f(\btheta)/g(\btheta)) = 1 + 2 \mbox{Var}_A(f(\btheta) /
  g_A(\btheta))
\end{equation}

\subsection{Importance Sampling Convergence}

With $f$, $g$, and $A$ as defined above, $f$ and $g_A$ have the same
domain.  Assuming errors in estimating $q$ and normalization errors
are of a lesser order of magnitude, we can treat the histogram heights
as being sampled from $g_A$.  Suppose the histogram has $l$ bins
$\{B_j\}$, each with width $h$ and based around the points
$g_A(\btheta_j)$.  Then by equation (\ref{imp var}),

\begin{eqnarray*}
  \mbox{Var}_A(f(\btheta) / g_A(\btheta)) &=&
     \sum_{j=1}^l \int_{B_j} \frac{(f(\btheta) - g_A(\btheta))^2}{g_A(\btheta)}d\btheta\\
     &=& \sum_{j=1}^l \int_{B_j} \frac{(g_A(\btheta) + \nabla g_A(\btheta)\cdot(\btheta_j - \btheta) + O((\btheta_j - \btheta)^2) - g_A(\btheta))^2}{g_A(\btheta)}d\btheta\\
     &=& \sum_{j=1}^l \int_{B_j} \frac{(\nabla g_A(\btheta)\cdot(\btheta_j - \btheta))^2 + O((\btheta_j - \btheta)^3)}{g_A(\btheta)}d\btheta\\
     &\leq& \sum_{j=1}^l \int_{B_j} \frac{\norm{\nabla \cdot g_A(\btheta)}^2 h^2}{g_A(\btheta)} d\btheta\\
     &=& h^2 \int \frac{\norm{\nabla \cdot g_A(\btheta)}^2}{g_A(\btheta)} d\btheta \label{h var}
\end{eqnarray*}

Because $h \propto m^{-d}$ where $d$ is the number of dimensions, and
$m$ is the number of samples used to make the histogram,

\begin{equation*}
  \mbox{Var}_A(f(\btheta) / g_A(\btheta)) \leq Cm^{-2/d}\\
\end{equation*}

\noindent where $C \propto \int \frac{\norm{\nabla \cdot
    g_A(\btheta)}^2}{g_A(\btheta)} d\btheta$.  Putting this together
with (\ref{I var}), we get

\begin{equation}
  \label{final variance}
  \mbox{Var}(I_n) = \mbox{Var}(p(\bx|T) / A_n) = n^{-1}(1+O(Cm^{-2/d}))
\end{equation}

\section{Implementation Issues}

\subsection{Speed of Convergence}

The variance of $n^{-1}(1+O(Cm^{-2/d}))$ given by (\ref{final
  variance}) is asymptotically equal to $n^{-1}$, which is the typical
importance sampling rate.  In practice however, the asymptotic results
cannot distinguish useful from impractical estimators.  If $Cm^{-2/d}$
is small and $\mbox{Var}(p(\bx|T) / A_n) \approx n^{-1}$, then
$p(\bx|T)$ can be approximated in only 1000 samples to about $6\% =
\frac{1.96}{\sqrt{1000}}$ with 95\% confidence.  For many theory
choice purposes, this is quite sufficient.  Thus in typical problem
cases the factor of $Cm^{-2/d}$ will be very significant.  If
$Cm^{-2/d} \gg 1$, then the convergence rate may in practice be
similar to $n^{-1}m^{-2/d}$.  Compare this to the rate of
$n^{-1}m^{-4/(4+d)}$ for the methods proposed by Zhang and
Neddermeyer.

This method also uses simple histograms, instead of a more
sophisticated density estimation method (Zhang uses kernel
estimation, Neddermeyer uses linear blend frequency polynomials).
Although simple histograms converge slower for large $d$ as shown
above, they are much faster to compute for large $d$.

Neddermeyer's LBFP algorithm is quite efficient compared to Zhang's,
but its running time is $O(2^dd^2n^\frac{d+5}{d+4})$.  $d$ is a
constant for any fixed problem, but if, say, $d=10$, then the
dimensionality constant multiplies the running time by $2^{10}10^2
\approx 10^5$.

By contrast, this paper's method takes only $O(dm\mbox{log}(m))$ time
to construct the initial histogram, and an additional
$O(dn\mbox{log}(m))$ time to do the importance sampling.  The main
reason for the difference is that querying a simple histogram can be
done in $\mbox{log}(m)$ time by computing the bin coordinates and
looking up the bin's height in a tree structure.  However, querying a
LBFP requires blending all nearby bins and is thus exponential in $d$.

\subsection{When $g = 0$}

Our discussion assumed that $g(\btheta) = p(\btheta|\bx, T)$ was
always positive.  If $g$ goes to 0 where the histogram is positive,
the variance of $A_n^{-1}$ will be infinite.  However, this paper's method
can still be used if $g(\btheta)$ is 0 over some well-defined area.

For instance, suppose one dimension $\theta_k$ of $p(\btheta|T)$ is
defined by a gamma distribution, so that $p(\theta_k|T) = 0$ if and
only if $\theta_k \leq 0$.  Then we can ensure the variance is not
infinite by checking that the histogram is only defined where
$\theta_k > \epsilon > 0$ for some fixed $\epsilon$.

The \texttt{margLikArrogance} package contains a simple mechanism to
do this.  The user may specify a range along each dimension of
$\btheta$ where it is known that $g > 0$.  If the histogram is
non-zero outside of this range, the method aborts with an error.

Note that the variance of the estimator increases with $\int
\frac{\norm{\nabla \cdot g_A(\btheta)}^2}{g_A(\btheta)} d\btheta$.  In
practice the estimator will work well only when $g$ doesn't go to 0
too quickly where the histogram is positive.  In these cases the
histogram will be defined well away from any region where $g=0$ and
infinite variance won't be an issue even if $g=0$ somewhere.

\subsection{Bin Shape}

Cubic histogram bins were used above---their widths were fixed at $h$
in each dimension.  Although the asymptotic results aren't affected by
the shape of each bin, for usable convergence rates the bins'
dimensions need to compatible with the shape of the high probability
region of $p(\btheta|\bx, T)$.  Unfortunately, it is difficult to
determine the best bin shapes.

The \texttt{margLikArrogance} package contains a simple workaround: by
default the distribution is first scaled so that the sampled standard
deviation along each dimension is constant.  This is equivalent to
setting each bin's width by dimension in proportion to that
dimension's standard deviation.  If this simple rule of thumb is
insufficient, the user can scale the sampled values of $p(\btheta|\bx,
T)$ manually (and make the corresponding adjustment to the estimate
$A_n$).

\section{Conclusion}

This paper has described an ``arrogance sampling'' technique for
computing the marginal likelihood or Bayes factor of a Bayesian model.
It involves using samples from the model's posterior parameter
distribution along with the scaled values of the distribution's
density at those points.  These samples are divided into two main
groups: $m$ samples are used to build a histogram; $n$ are used to
importance sample the histogram using the posterior parameter
distribution as the proposal.

This method is simple to implement and runs quickly in
$O(d(m+n)\mbox{log}(m))$ time.  Its asymptotic convergence rate,
$n^{-1}(1+O(Cm^{-2/d}))$, is not remarkable, but in practice
convergence is fast for many problems.  Because the required inputs
are similar to those of the harmonic mean estimator, it may be a
convenient replacement for it.

\section{References}

\begin{enumerate}
  \item S. Chib.  ``Marginal Likelihood from the Gibbs Output''
    \emph{Journal of the American Statistical Association}.  Vol 90,
    No 432.  (1995)

  \item S. Chib and I. Jeliazkov.  ``Accept-reject Metropolis-Hastings
    sampling and marginal likelihood estimation'' \emph{Statistica
      Neerlandica}. Vol 59, No 1.  (2005)
  
  \item A. Gelman and X. Meng.  ``Simulating Normalizing Constants:
    From Importance Sampling to Bridge Sampling to Path Sampling''
    \emph{Statistical Science}. Vol 13, No 2. (1998)

  \item R. Kass and A. Raftery.  ``Bayes Factors'' \emph{Journal of the
  American Statistical Association}.  Vol 90, No 430.  (1995)

  \item R. Neal.  ``The Harmonic Mean of the Likelihood: Worst Monte
    Carlo Method Ever''.  Blog post,
    \texttt{http://radfordneal.wordpress.com/2008/08/17/the-harmonic-mean-of-the-likelihood-worst-monte-carlo-method-ever/}.  (2008)
    
  \item J. Neddermeyer.  ``Computationally Efficient Nonparametric
    Importance Sampling'' \emph{Journal of the American Statistical
      Association}.  Vol 104, No 486.  (2009)  	arXiv:0805.3591v2
  
  \item A. Owen and Y. Zhou. ``Safe and effective importance
    sampling'' \emph{Journal of the American Statistical
      Association}. Vol 95, No 449.  (2000)

  \item C. Robert and D. Wraith.  ``Computational methods for Bayesian
    model choice'' arXiv:0907.5123v1

  \item P. Zhang.  ``Nonparametric Importance Sampling'' \emph{Journal
    of the American Statistical Association}. Vol 91, No 435. (1996)

  \item M. Zlochin and Y. Baram.  ``Efficient Nonparametric
    Importance Sampling for Bayesian Inference'' \emph{Proceedings
      of the 2002 International Joint Conference on Neural Networks}
    2498--2502.  (2002)

\end{enumerate}

\end{document}